# An Updated Ultraviolet Calibration for the Swift/UVOT

A. A. Breeveld[a], W. Landsman[b], S. T. Holland[b,c,d], P. Roming[e], N. P. M. Kuin[a] and M. J. Page[a]

[a]*Mullard Space Science Laboratory, University College London, Holmbury St. Mary, Dorking, Surrey RH5 6NT, UK*
[b]*NASA/Goddard Space Flight Center, Greenbelt, MD 20771*
[c]*Universities Space Research Association, 10211 Wincopin Circle, Suite 500, Columbia, MD 21044*
[d]*Center for Research and Exploration in Space Science and Technology, NASA/GSFC, Greenbelt, MD20771*
[e]*Southwest Research Institute, Department of Space Science,6220 Culebra Rd, San Antonio, TX 78238*

**Abstract.** We present an updated calibration of the Swift/UVOT broadband ultraviolet (*uvw1*, *uvm2*, and *uvw2*) filters. The new calibration accounts for the ~1% per year decline in the UVOT sensitivity observed in all filters, and makes use of additional calibration sources with a wider range of colours and with HST spectrophotometry. In this paper we present the new effective area curves and instrumental photometric zeropoints and compare with the previous calibration.

## INTRODUCTION

*Swift* [1] was launched in late 2004 to observe Gamma Ray Bursts (GRB) and other transient or variable phenomena, with three complementary and parallel instruments on board: the Ultraviolet and Optical Telescope (UVOT), X-ray telescope ([2]) and the Burst Alert Telescope ([3]). The UVOT is sensitive within the range 1600 to 8000Å and has a *white* light filter to sample the whole wavelength range, or a choice of 6 lenticular filters to select bands in the optical (*v*, *b* and *u*) and ultraviolet (*uvw1*, *uvm2* and *uvw2*) ranges. There are also two grisms for low resolution spectroscopy. Details about the instrument design and performance can be found in [4] and [5].

The in-orbit photometric calibration given in [5] was based on the ground calibration of the telescope throughput and filter effective areas, adjusting these to correctly predict the count rates measured from standard stars observed at the start of the mission. In the last few years we have been able to measure a decline in sensitivity at the rate of ~1% per year in all filters, and a correction is now available for this at the HEASARC.

Using additional calibration sources with a wide range of colours we have found that for the ultraviolet filters only, the calibration given in [5] over-predicted the count rates for red stars. We have therefore modified the effective area curves for the *uvm2* and *uvw2* filters slightly at the red end. The *uvw1* effective area curve required a more significant change in order to give consistent photometry of both hot and cool stars. The calibration of the effective area curves has been further improved by increasing the number and colour range of standard stars used, and also by using more observations for each standard star.

New zeropoints have been derived for the UVOT instrumental magnitude system using these revised effective area curves, with the benefit of many additional observations and a more up-to-date spectrum of Vega, on which our magnitude system is based. All other photometric products have also been updated in line with these new zeropoints and effective areas.

## TIME-DEPENDENT SENSITIVITY CORRECTION

Photometric standard sources have been observed in all filters regularly throughout the mission, to monitor any changes in count rate which would point to a change in the instrument throughput. All the data from these 16 sources throughout the mission have recently been reanalyzed with the most up-to-date software, including the large scale sensitivity (LSS) correction described in [6]. The count rates, measured with the standard 5 arcsec aperture, were normalized using the mean count rate for each star in each filter in exposures taken within the first 500 days to allow all the stars to be plotted and fitted together. In each case the data were fitted with a weighted straight line fit. The slopes for all filters except the white are similar to each other, and are consistent with a decrease of 1% per year. The *b* filter is shown in Fig. 1 as an example. It turned out that some of the additional scatter in the *white* filter was caused by high background count rates and a simple fit to correct for the background brings the *white* filter also within the range of 1% decline per year.

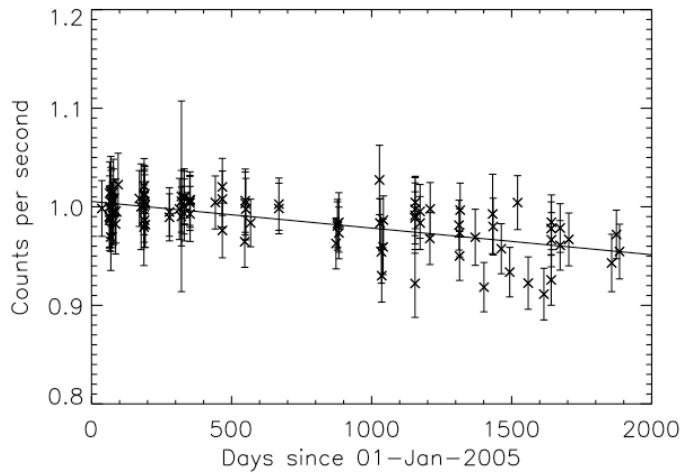

**FIGURE 1.** Count rates of standard stars in the *b* filter, normalized to the count rates within the first 500 days post launch, against days since launch. The straight-line fit corresponds to a decline of 0.99±0.18 % per year.

## CHANGES TO THE EFFECTIVE AREA CURVES

The original in-orbit photometry in the ultraviolet depended on only three white dwarf standard stars observed at the start of the mission. Once we had the time-dependent sensitivity correction in place we were able to include many observations of the three white dwarfs taken throughout the mission, and we also included more standard stars with a wider range of colours (see Table 1). Count rates for all the stars were obtained using UVOTSOURCE, which is part of the suite of Swift software provided by the Swift Science Center[1]. UVOTSOURCE performs all the necessary corrections including corrections for LSS ([6]) and time-dependent sensitivity.

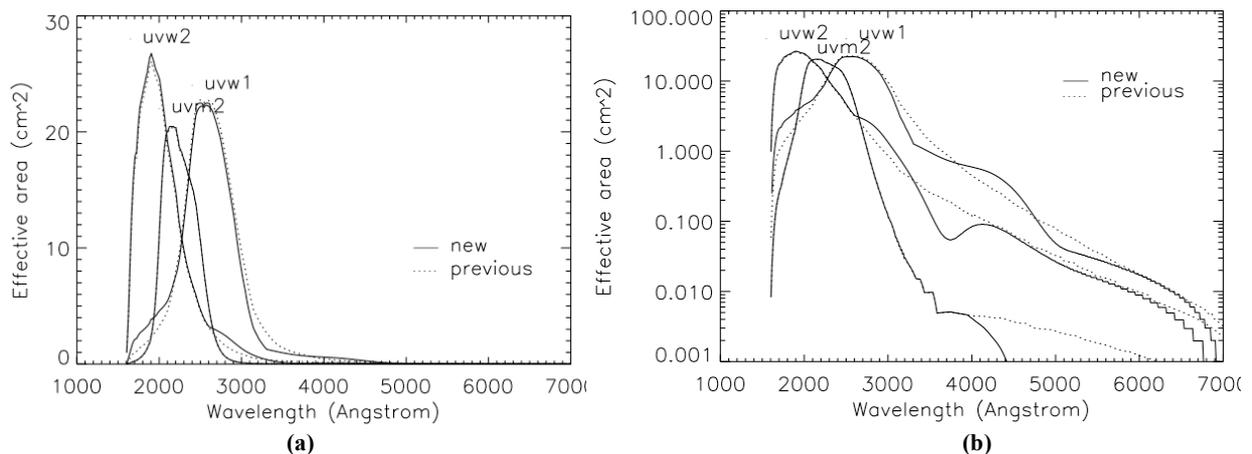

**FIGURE 2.** Comparing in-orbit effective area curves from [5] (dotted lines) with the new ones from this version (solid lines). The plots are the same except that (b) uses a log scale to show red tails more clearly.

---

[1] http://swift.gsfc.nasa.gov/docs/swift/swiftsc.html

The expected count rate of each observed standard star for each filter was calculated by convolving the known spectrum (see second column of Table 1) of the observed star with the effective area curve. Comparison of the expected with the measured count rates for the range of stars of different colours showed that there was a problem with the shape of the effective area curves, particularly at the red end for *uvm2* and *uvw2* and more generally for *uvw1*. To minimize changes from the ground-based measurements, the *uvm2* filter was only modified longward of 3950 Å, and the *uvw2* filter only modified longward of 2600 Å. The *uvw2* filter also required a tiny contribution at long (>7000 A) wavelengths to predict the photometry of the very red stars. However, the *uvw1* filter required modifications across the entire wavelength range in order to match the photometry of stars of all colours. Figure 2 shows a comparison of the ultraviolet effective areas given in [5] and the new effective areas.

For each source in each filter we calculated a ratio of measured count rate to expected count rate. Excluding the reddest stars, the ratios were averaged over each filter and these average ratios were used to renormalise the effective area curves by a small factor.

**TABLE (1).** The sources used for the effective area analysis and photometric zeropoint calculation. The second column gives the origin of the calibrated spectrum. The CALSPEC spectra were obtained from ftp://ftp.stsci.edu/cdbs/current_calspec/; NGSL from http://archive.stsci.edu/prepds/stisngsl/; IUE from Holberg et al. (2003). The spectrum of MMJ6476 has a model atmosphere added below 1700Å. The HD118055 spectrum has had a stellar radial velocity of -101 km/s added.
*these sources were only used for checking the red response and were not included in the zeropoint calculation.

| Object name | Origin | V mag | B mag | Type |
|---|---|---|---|---|
| WD1121+145 | IUE | 16.9 | 16.6 | DA |
| WD1026+453 | CALSPEC | 16.1 | 15.9 | DA |
| WD1657+343 | CALSPEC | 16.4 | 16.2 | DA |
| LDS749B | CALSPEC | 14.67 | 14.71 | DBQ4 |
| HZ4 | CALSPEC | 14.51 | 14.6 | DA4 |
| MMJ6476 | NGSL | 10.9 | 11.5 | A7.2:m |
| P041C | CALSPEC | 12.00 | 12.62 | G0V |
| P330E | CALSPEC | 12.92 | 13.64 | G0V |
| P177D | CALSPEC | 13.47 | 14.13 | G0V |
| HD118055* | NGSL | 8.89 | 10.1 | K0w |
| BD+442051* | NGSL | 8.68 | 10.22 | M2V |
| HD102780* | NGSL | 8.18 | 9.8 | K2 |

## PHOTOMETRIC ZEROPOINTS

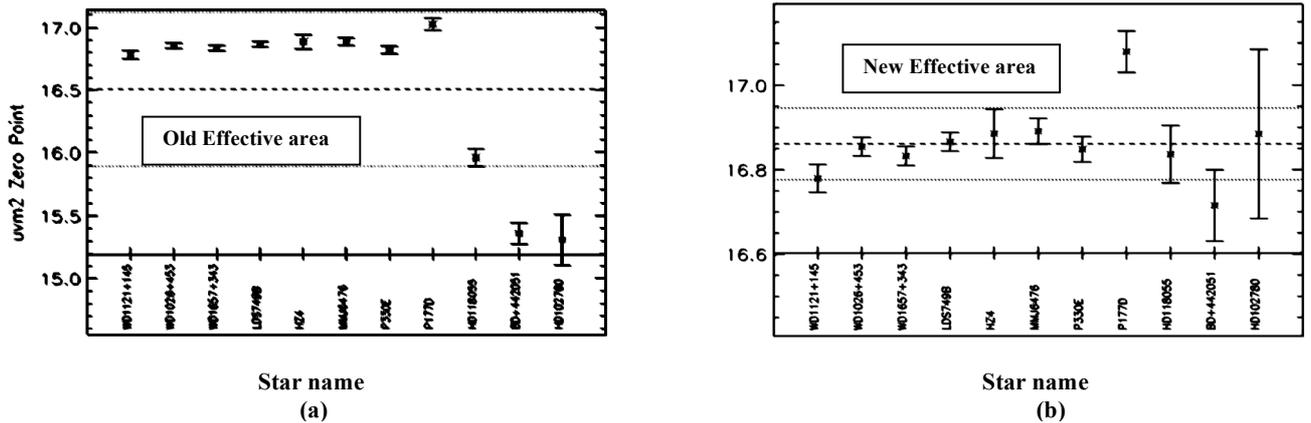

**FIGURE 3.** An example of how the new effective area has improved for redder stars for *uvm2*. Each data point represents the zero point as calculated according to the measurement of one standard star. The average zeropoint is shown with a dashed line and the RMS scatter is indicated by the two dotted lines. Error bars on the points include the Poisson error in the raw count rate as well as error associated with the stellar specrum (a) Predicted count rates using effective areas from [5]; the stars are in colour order (blue to red). Note that the three reddest stars are out by more than a magnitude (and would lower the zeropoint too much). (b) Using new effective areas. Now the zeropoints based on red and blue stars are in agreement.

The UVOT photometric calibration is based on the standard star Vega. For this zeropoint calculation we introduced a newer spectrum of Vega[2].

For each filter, Zeropoints were calculated by convolving the spectrum of Vega (defined in the UV as having a magnitude of zero) with the effective area, modified by the ratio of observed to expected count rates for each star (see [5] for more details). Figure 3 shows the zeropoint values calculated for the *uvm2* filter for the whole range of standard stars. The final zeropoints (given in Table 2) are an average of those obtained from individual stars, but leaving out the red stars (the last three in Fig. 3).

All the other calibration products in the CALDB[1] that depend on the photometry (count to flux ratio, galactic and zodiacal background light predictions and the AB magnitude system) and the webtool[2] have been updated with the new effective areas and zeropoints.

**TABLE (2).** The new zeropoints, and for comparision the previous set of zeropoints given in [5]. The third column gives the new AB zeropoints. The optical and *white* zeropoints are included here for convenience although they have not changed.

| Filter | New Zeropoints (Vega) | Previous Zeropoints (Vega) | New AB Zeropoints |
|---|---|---|---|
| V | 17.89 ± 0.01 | 17.89 ± 0.01 | 17.88 ± 0.01 |
| B | 19.11 ± 0.02 | 19.11 ± 0.02 | 18.98 ± 0.02 |
| U | 18.34 ± 0.02 | 18.34 ± 0.02 | 19.36 ± 0.02 |
| uvw1 | 17.44 ± 0.03 | 17.49 ± 0.03 | 18.95 ± 0.03 |
| uvm2 | 16.85 ± 0.03 | 16.82 ± 0.03 | 18.54 ± 0.03 |
| uvw2 | 17.38 ± 0.03 | 17.35 ± 0.03 | 19.11 ± 0.03 |
| White | 20.29 ± 0.04 | 20.29 ± 0.04 | 21.09 ± 0.04 |

## CONCLUSIONS

We have presented a new calibration for the UVOT UV filters (*uvw1*, *uvm2* and *uvw2*). The optical (*u, b, v*) and *white* calibration has not changed. Using the new UV effective area curves improves the understanding of the UVOT UV response to sources of a wide range of colours. The new calibration also accounts for the gradual decline in throughput observed since launch in all filters.

We have provided new zeropoints consistent with the effective areas. The difference from the previous release is less than 0.05 magnitudes. The updated CALDB[3] was released at the end of 2010 with an update to the new effective area curves and zeropoints as well as all related photometry products such as count to flux conversions.

The webtool[4] for predicting count rates has also been updated to use the new calibration. This should now give a more accurate prediction for expected count rates of red stars.

## ACKNOWLEDGMENTS

This work is supported at MSSL by funding from STFC and at PSU by NASA's Office of Space Science through grant NAG5-8401 and NAS5-00136. We acknowledge the use of public data from the Swift archive. We would like to thank Tracey Poole for her IDL routines.

## REFERENCES

1. N. Gehrels et al., ApJ, 2004, 611, 1005
2. D. N. Burrows et al., Space Sci. rev., 2005, 120, 165
3. S. D. Barthelmy et al., Space Sci. Rev., 2005, 120, 143
4. P. W. A. Roming et al., Space Sci. Rev., 2005, 120, 95
5. T. S. Poole et al., MNRAS, 2008, 383, 627
6. A. A. Breeveld et al., MNRAS, 2010, 406, 1687

---

[2] ftp://ftp.stsci.edu/cdbs/current_calspec/alpha_lyr_stis_005.fits
[3] CALDB can be found at: http://swift/gsfc.nasa.gov/docs/heasarc/caldb/swift
[4] The Webtool gives predicted count rates for given star types and can be found at: http://www.mssl.ucl.ac.uk/www_astro/uvot/